\documentclass[english,aps,prl,reprint,superscriptaddress]{revtex4-1}
\usepackage[T1]{fontenc}
\usepackage[latin9]{inputenc}
\usepackage{color}
\usepackage{babel}
\usepackage{amsmath}
\usepackage{amssymb}
\usepackage{graphicx}
\usepackage[unicode=true,
 bookmarks=true,bookmarksnumbered=false,bookmarksopen=false,
 breaklinks=false,pdfborder={0 0 0},pdfborderstyle={},backref=false,colorlinks=true]
 {hyperref}
\hypersetup{
 allcolors=blue}

\makeatletter

\newcommand*\LyXThinSpace{\,\hspace{0pt}}

\usepackage{braket}
\usepackage{dsfont}

\makeatother

\begin{document}

\title{Role of quantum coherence in the thermodynamics of energy transfer }

\author{Ivan Henao}
\email{ivan.henao@ufabc.edu.br}

\selectlanguage{english}%

\affiliation{Centro de Ciências Naturais e Humanas, Universidade Federal do ABC,
Avenida dos Estados 5001, 09210-580 Santo André, São Paulo, Brazil}

\author{Roberto M. Serra}
\email{serra@ufabc.edu.br}

\selectlanguage{english}%

\affiliation{Centro de Ciências Naturais e Humanas, Universidade Federal do ABC,
Avenida dos Estados 5001, 09210-580 Santo André, São Paulo, Brazil}

\affiliation{Department of Physics, University of York, York YO10 5DD, United
Kingdom}
\begin{abstract}
Recent research on the thermodynamic arrow of time, at the microscopic
scale, has questioned the universality of its direction. Theoretical
studies showed that quantum correlations can be used to revert the
natural heat flow (from the hot body to the cold one), posing an apparent
challenge to the second law of thermodynamics. Such an ``anomalous''
heat current was observed in a recent experiment (arXiv:1711.03323),
by employing two spin systems initially quantum correlated. Nevertheless,
the precise relationship between this intriguing phenomenon and the
initial conditions that allow it is not fully evident. Here, we address
energy transfer in a wider perspective, identifying a nonclassical
contribution that applies to the reversion of the heat flow as well
as to more general forms of energy exchange. We derive three theorems
that describe the energy transfer between two microscopic systems,
for arbitrary initial bipartite states. Using these theorems, we obtain
an analytical bound showing that certain type of quantum coherence
can optimize such a process, outperforming incoherent states. This
genuine quantum advantage is corroborated through a characterization
of the energy transfer between two qubits. For this system, it is
shown that a large enough amount of coherence is necessary and sufficient
to revert the thermodynamic arrow of time. As a second crucial consequence
of the presented theorems, we introduce a class of nonequilibrium
states that only allow unidirectional energy flow. In this way, we
broaden the set where the standard Clausius statement of the second
law applies. 
\end{abstract}
\maketitle
Since Carnot discovered the fundamental limit that governs the efficiency
of heat engines, the second law of thermodynamics has been discussed
and explored in different ways. One of them refers to the celebrated
Clausius statement that heat must flow from a hot system to a cold
one, when the whole system is isolated. This preferred direction of
the heat flux may be interpreted as a ``thermodynamic arrow'' that
characterizes the time ordering of physical events \cite{1-Boltzmann-time-arrow,2-phys-bas.-dir-time,3-Eddington-arrow-of-time}.
More recently, developments on quantum thermodynamics have allowed
the thermodynamic description of microscopic quantum systems. Fluctuation
relations \cite{4-review-FTs,5-colloquium-FTs} and information-theory
inspired approaches \cite{6-fund.paper-RT,7-RT-of-out-of-eq-states,8-Fund-lim,Winter-2017}
represent powerful tools to carry out this task. These new paradigms,
which refer to systems that start in a nonequilibrium state or that
undergo a nonequilibrium dynamics, have led to some generalizations
of the second law beyond the scope of standard thermodynamics \cite{10-Jarzynski,11-Crooks,12-second-laws,13-coh+free-energy}.
They also establish new connections between thermodynamics and information
theory \cite{14-therm-of-inf,15-role-of-inf-in-QT}, enabling a formal
treatment of Maxwell's demon and related subjects \cite{16-colloquium-max-dem,17-therm-of-inf-proc,18-PCamati,19-Lutz-Ciliberto}.
On the experimental side, crucial advances have been achieved to access
and characterize energy fluctuations in microscopic systems \cite{20-(Exp-Jarzynski-science),21-(inf-to-energy-conversion-nature),22-(Jarz-electronic),23-(Jarz-traped-ions),24-(Tiago),25-(MD-exp-1),26-(MD-exp-2),27-(phot-MD),28-(inf-to-energy-conver),29-Huard-2017,key-1 (Marcela)}. 

Among the plethora of results obtained on quantum thermodynamics,
many of them center on the concept of work \cite{30-Quant.Therm.,31-work-from-op-princ},
and its interplay with other thermodynamic variables such as entropy
production \cite{32-Entr-prod}. The process of heat exchange between
two finite-size systems is a less studied phenomenon. While for two
quantum systems in an initially uncorrelated state the Clausius statement
holds \cite{33-Partovi}, the same may not be true for initially correlated
bodies \cite{33-Partovi,34-ent+therm-arrow}. This assertion has been
corroborated in a recent experiment using a two-spin system embedded
in a Nuclear Magnetic Resonance setup \cite{35-exp-rev-of-arrow}.
However, a comprehensive description of such a behavior is lacking.
Of particular interest is to unveil the role played by quantum properties,
e.g. coherence or entanglement, in the reversion of the thermodynamic
arrow of time. In this respect, a fluctuation relation for heat exchange
in the presence of classical correlations was derived and discussed
in Ref.~\cite{36-FT-for-correl-systems}. \textcolor{black}{The performance
of} quantum coherence has been analyzed in the context of work extraction
\cite{37-workfromcoh,38-manycyclemachines,39-coh+meas,40-Kosloff,41-colloqcoh},
entropy production \cite{42-Plastina,43-Landi}, and Landauer's erasure
\cite{44-Campbell}. Other investigations focus on how coherence transforms
under thermodynamic operations \cite{45-Qcoh-timetr-therm,46-lim-on-evol-of-QC,47-energycost-of-coh},
without intending to assign it some operational meaning. 

We investigate the physical process of energy exchange between two
microscopic systems,\textcolor{red}{{} }\textcolor{black}{consistently}
with the first law of thermodynamics \cite{key-2 Petruccione}. To
this aim, we present three theorems that describe the transfer of
average energy induced by unitary and energy-conserving evolutions.
These theorems are valid for arbitrary bipartite states, allowing
to incorporate heat exchange (i.e., when each system starts at thermal
equilibrium) as a particular case. Two fundamental consequences are
derived from such theorems. The first one establishes that quantum
coherence (in the eigenbasis of the free joint Hamiltonian) \textcolor{black}{may
enhance} the energy transfer, under an optimal evolution. Specifically,
it is shown that the maximum energy transfer for an incoherent initial
state is upper bounded by the obtained one when coherence of certain
type is included. Next, we deduce a class of states that restrict
the energy flow to a single direction. This set contains all the tensor
products between thermal states, in agreement with the Clausius statement
of the second law, but is not restricted to them. Finally, we apply
our general results to the two-qubit example. We verify that, for
an optimal dynamics, the energy transfer is maximized only if enough
initial ``useful'' coherence is available. This quantum feature
is further responsible for reverting the heat flow between thermal
qubits, which provides a comprehensive framework for the experimental
results reported in \cite{35-exp-rev-of-arrow}. \textcolor{black}{We
also obtain a linear relation between the maximum energy transfer
and the concurrence, for entangled Bell-diagonal states \cite{48-belldiag1,49-belldiag2}.
It is shown that a quantum enhancement results from a subset of separable
states within this class.}\textcolor{red}{{} }

\textit{Energy exchange under SEC unitaries}.\textemdash{} Let us
consider two quantum systems $A$ and $B$, with nondegenerate and
discrete Hamiltonians $H_{A}$ and $H_{B}$, respectively. We adopt
``energy conservation'' according to the condition $[U,H]=0$, where
$H=H_{A}+H_{B}$ is the total \textit{free} Hamiltonian and $U$ is
a unitary map generated by some interaction Hamiltonian $H_{I}$.
This is equivalent to demand that $U$ preserves the sum of the local
energies for any initial joint state $\rho$: $\textrm{Tr}\left(HU\rho U^{\dagger}\right)=\textrm{Tr}\left(H\rho\right)$.
Hence, we say that $U$ is ``Strong Energy Conserving'' (SEC). For
small systems, the strength of the interaction energy may be of the
same order of the local energies. Therefore, it is not evident how
to physically implement $U$, as even for $H_{I}$ constant we can
only guarantee that the total energy (including the contribution from
$H_{I}$) is preserved. \textcolor{black}{A sufficient condition for}
$U$ to be SEC comes from the relation $[H,H_{I}]=0$ \cite{50-comment about SEC U}.
The resonant Jaynes-Cummings model \cite{51-Quantum Optics} exemplifies
a well known system that fulfills this requirement. If, in addition,
we assume that the total Hamiltonian at the beginning and at the end
of the energy exchange process is $H$, the local energies become
well defined quantities. It is worth remarking that the adopted definition
constitutes a paradigmatic approach to the first law of thermodynamics
in microscopic systems (see e.g. \cite{15-role-of-inf-in-QT,30-Quant.Therm.}
and references therein). 

Without loss of generality, the state $\rho$ can be written as 
\begin{equation}
\rho=\rho_{Diag}+\chi,\label{eq:1 (decomp. of rho)}
\end{equation}
where $\rho_{Diag}$ and $\chi$ are the diagonal part and the coherent
(off-diagonal) part of ${\normalcolor {\color{red}{\normalcolor \rho}}}$
in the eigenbasis of $H$, respectively.\textcolor{red}{{} }Each eigenstate
of $H$ with eigenvalue $E$\textbf{ }has the form $\left|i_{E}\right\rangle _{A}\left|j_{E}\right\rangle _{B}$,
where $\left|i_{E}\right\rangle _{A}$ and $\left|j_{E}\right\rangle _{B}$
are local energy eigenstates satisfying $H_{A}\left|i_{E}\right\rangle _{A}=\varepsilon_{i_{E}}\left|i_{E}\right\rangle _{A}$,
$H_{B}\left|j_{E}\right\rangle _{B}=\bar{\varepsilon}_{j_{E}}\left|j_{E}\right\rangle _{B}$,
and $\varepsilon_{i_{E}}+\bar{\varepsilon}_{j_{E}}=E$. For $E$ fixed,
the relation $\varepsilon_{i_{E}}+\bar{\varepsilon}_{j_{E}}=E$ and
the non-degeneracy of the local Hamiltonians imply a one-to-one correspondence
between $i_{E}$ and $j_{E}$. This means that for each $i_{E}$ there
is only one $j_{E}$ that fulfills this equation and viceversa. In
this way, we can completely characterize the spectrum of $H$ by using
the total energy index $(E)$ and a single local energy index. Choosing
by convention the index $i_{E}$, the resulting set is denoted as
$\{\left|i_{E},E\right\rangle \}_{i_{E},E}$, where $H\left|i_{E},E\right\rangle =E\left|i_{E},E\right\rangle $
and $H_{A}\left|i_{E},E\right\rangle =\varepsilon_{i_{E}}\left|i_{E},E\right\rangle $.
This notation provides a natural decomposition into subspaces of fixed
energy $E$, very suitable for the analysis of SEC unitaries. 

\textcolor{black}{Now we explicitly write $\rho_{Diag}$ and $\chi$
in the eigenbasis} $\left\{ \left|i_{E},E\right\rangle \right\} $.
For $\rho_{Diag}$ we have:
\begin{equation}
\rho_{Diag}=\sum_{E}p_{E}\rho_{Diag}(E),\label{eq:2 (rho_diag)}
\end{equation}
where $\rho_{Diag}(E)\equiv\sum_{i_{E}}\frac{p(\varepsilon_{i_{E}},E)}{p_{E}}\Pi_{i_{E}}^{(E)}$
and $\Pi_{i_{E}}^{(E)}\equiv\left|i_{E},E\right\rangle \left\langle i_{E},E\right|$.
The joint probability to measure energy $\varepsilon_{i_{E}}$ for
system $A$ and total energy $E$ is given by $p(\varepsilon_{i_{E}},E)$.
Accordingly, $p_{E}=\sum_{i_{E}}p\left(\varepsilon_{i_{E}},E\right)$
is the total probability to measure joint energy equal to $E$. On
the other hand, 
\begin{equation}
\chi=\sum_{E,E'}\chi(E,E'),\label{eq:3 (rho_coh)}
\end{equation}
where ${\normalcolor \chi(E,E')\equiv\sum_{i_{E},j_{E'}:\,i_{E}\neq j_{E'}\textrm{ or }E\neq E'}{\color{red}{\normalcolor \alpha_{i_{E},j_{E'}}^{(E,E')}}}\Pi_{i_{E},j_{E'}}^{(E,E')}}$
and $\Pi_{i_{E},j_{E'}}^{(E,E')}\equiv\left|i_{E},E\right\rangle \left\langle j_{E'},E'\right|$
\cite{52-comment-about-coh}. 

The energy transfer to the system $\gamma$ ($\gamma=A,B$) is denoted
as $\Delta\bigl\langle H_{\gamma}\bigr\rangle$ and represents the
average energy variation undergone by this system, through the application
of a SEC unitary $U$. If the initial joint state is $\rho$, then
$\Delta\bigl\langle H_{\gamma}\bigr\rangle={\color{red}{\normalcolor \textrm{Tr}\left(H_{\gamma}U\rho U^{\dagger}\right)}}-\textrm{Tr}\left(H_{\gamma}\rho\right)$.
Taking into account Eq.~(\ref{eq:1 (decomp. of rho)}), this quantity
is given by 
\begin{equation}
\Delta\bigl\langle H_{\gamma}\bigr\rangle=\Delta_{Diag}\bigl\langle H_{\gamma}\bigr\rangle+\Delta_{Coh}\bigl\langle H_{\gamma}\bigr\rangle,\label{eq:4 (decomp of energy transfer)}
\end{equation}
where $\Delta_{Diag}\bigl\langle H_{\gamma}\bigr\rangle\equiv{\color{red}{\normalcolor \textrm{Tr}\left(H_{\gamma}U\rho_{Diag}U^{\dagger}\right)}}-\textrm{Tr}\left(H_{\gamma}\rho_{Diag}\right)$
and $\Delta_{Coh}\bigl\langle H_{\gamma}\bigr\rangle\equiv{\color{red}{\normalcolor \textrm{Tr}\left(H_{\gamma}U\chi U^{\dagger}\right)}}-\textrm{Tr}\left(H_{\gamma}\chi\right)$
are the ``diagonal energy transfer'' and the ``coherent energy
transfer'', respectively. 

Definitions 1-3 set the framework for the presentation of Theorems
1-3. These theorems characterize the energy transfer for arbitrary
initial states and are pivotal in the derivation of subsequent results.
We leave the corresponding proofs to the Supplemental Material \cite{53-Supplementary-Material},
in order to focus on their physical aspect. 

\textbf{\textcolor{black}{Definition~1}}. From the eigenspace $\mathcal{H}_{E}\equiv\textrm{span}\left\{ \left|i_{E},E\right\rangle \right\} _{i_{E}}$,
spanned by all the joint eigenstates with eigenenergy $E$, we introduce
the $E$-local subspace of system $A$, $\mathcal{H}_{E}^{A}\equiv\textrm{span}\left\{ \left|i_{E}\right\rangle _{A}\right\} $.
A state $\varrho^{A}(E)$ with eigenvectors $\left|\varphi_{i}^{E}\right\rangle _{A}\in\mathcal{H}_{E}^{A}$
is called an $E$-local state of system $A$. In addition, an $E$-local
unitary $U_{E}^{A}$ is a unitary that maps the subspace $\mathcal{H}_{E}^{A}$
in\textcolor{black}{to itself,} and is exclusively defined on this
subspace. 

\textbf{\textcolor{black}{Definition~2}}. The requirement that a
SEC $U$ preserves the total energy for any state $\rho$ is equivalent
to demand that it does so for any joint energy eigenstate $\left|i_{E},E\right\rangle $.
That is, $U$ must transform $\left|i_{E},E\right\rangle $ into a
superposition of eigenstates having equal total energies: $U\left|i_{E},E\right\rangle \equiv\sum_{j_{E}}c_{i_{E},j_{E}}^{(E)}\left|j_{E},E\right\rangle $.
Taking into account that the action of $U$ is arbitrary within each
eigenspace $\mathcal{H}_{E}$, for any $E$ the coefficients $\{c_{i_{E},j_{E}}^{(E)}\}$
allow to construct an arbitrary $E$-local unitary: $U_{E}^{A}|i_{E}\rangle_{A}\equiv{\normalcolor {\color{green}{\normalcolor \sum}_{{\normalcolor j_{E}}}}}c_{i_{E},j_{E}}^{(E)}|j_{E}\rangle_{A}$. 

\textbf{\textcolor{black}{Definition~3}} \textcolor{black}{(restricted
passivity)}. Any $E$-local state $\varrho^{A}(E)=\sum_{i}q_{i}\left|\varphi_{i}^{E}\right\rangle _{A}\langle\varphi_{i}^{E}|$
can be transformed by an $E$-local unitary in the state $\varrho_{P}^{A}(E)=\sum_{i_{E}}q_{i_{E}}\left|i_{E}\right\rangle _{A}\langle i_{E}|$,
where $q_{i_{E}}\geq q_{i_{E}'}$ implies that $\varepsilon_{i_{E}}<\varepsilon_{i'_{E}}$,
for any $i_{E},i'_{E}$. We say that $\varrho_{P}^{A}(E)$ is $E$-passive,
or passive \textit{within} $\mathcal{H}_{E}^{A}$. Physically, this
means that $\varrho_{P}^{A}(E)$ is the state of minimum energy, that
can be attained from $\varrho^{A}(E)$ through an $E$-local unitary.
Analogously, the maximum energy state that results from applying an
$E$-local unitary on $\varrho^{A}(E)$ is: $\varrho_{M}^{A}(E)=\sum_{i_{E}}q_{i_{E}}\left|i_{E}\right\rangle _{A}\langle i_{E}|$,
such that $q_{i_{E}}\geq q_{i_{E}'}$ implies $\varepsilon_{i_{E}}>\varepsilon_{i'_{E}}$,
for any $i_{E},i'_{E}$ \cite{54-commentpassivity}. 

\textbf{\textcolor{black}{Theorem~1}}. Let $\rho_{Diag}^{A}(E)\equiv\textrm{Tr}_{B}\rho_{Diag}(E)$
be an $E$-local state defined through Eq. (\ref{eq:2 (rho_diag)}).
Under the effect of a SEC unitary $U$, the diagonal energy transfer
to system $A$ is given by $\Delta_{Diag}\bigl\langle H_{A}\bigr\rangle=\sum_{E}p_{E}\textrm{Tr}_{A}\left(H_{A}U_{E}^{A}\rho_{Diag}^{A}(E)U_{E}^{A\dagger}-H_{A}\rho_{Diag}^{A}(E)\right)$. 

\textit{Physical relevance}: This theorem allows us to straightforwardly
establish the possible values for the diagonal energy transfer. Since
any $E$-local unitary $U_{E}^{A}$ is arbitrary on $\mathcal{H}_{E}^{A}$,
according to Definition 2, the minimum (maximum) of $\Delta_{Diag}\bigl\langle H_{A}\bigr\rangle$
is determined by separately minimizing (maximizing) each term $\textrm{Tr}_{A}\left(H_{A}U_{E}^{A}\rho_{Diag}^{A}(E)U_{E}^{A\dagger}\right)$
with respect to $U_{E}^{A}$. From Definition 3, these extremal values
are attained when $U_{E}^{A}\rho_{Diag}^{A}(E)U_{E}^{A\dagger}$ is
an $E$-passive state (minimum), or a maximum energy $E$-local state
(maximum). The corresponding optimal SEC $U$ is readily obtained
by means of Definition 2. Theorem 1 is also fundamental for the proof
of Theorem 3 \cite{53-Supplementary-Material}. 

\textbf{\textcolor{black}{Theorem~2}}. Under the effect of a SEC
unitary $U$, the coherent energy transfer to system $A$ is given
by $\Delta_{Coh}\bigl\langle H_{A}\bigr\rangle=\sum_{k}\eta_{k}\varepsilon_{k}$,
where $\{\varepsilon_{k}\}$ are the eigenvalues of $H_{A}$ and $\eta_{k}\equiv2\sum'_{E}\sum_{i_{E}<j_{E}}\textrm{Re}\left(\alpha_{i_{E},j_{E}}^{(E,E)}c_{i_{E},k}^{(E)}c_{j_{E},k}^{(E)\ast}\right)$.
For $k$ fixed, the sum $\sum'_{E}$ is restricted to values of $E$
satisfying $E=\varepsilon_{k}+\bar{\varepsilon}_{l_{E}}$, where $\bar{\varepsilon}_{l_{E}}$
is an eigenvalue of $H_{B}$ (this implies $k\in\{k_{E}\}$). 

\textit{Physical relevance}: The coefficients $\eta_{k}$ embody an
interplay between the coefficients of coherence, $\alpha_{i_{E},j_{E}}^{(E,E)}$,
and the $c_{i_{E},k_{E}}^{(E)}$, which describe the action of $U$.
In particular, they are independent of $\alpha_{i_{E},j_{E'}}^{(E,E')}$,
for $E\neq E'$. In this way, this theorem singles out the kind of
coherence that may contribute to the energy transfer, corresponding
to those terms $\chi(E,E')$ with $E=E'$ in Eq. (\ref{eq:3 (rho_coh)}).
The corollary below states a necessary condition on $U$ to get a
non null coherent energy transfer. 

\textbf{\textcolor{black}{Corollary~2.1}}. Any SEC unitary with the
potential to yield $\Delta_{Coh}\bigl\langle H_{A}\bigr\rangle\neq0$
must belong to the following set: $\{\mathcal{U}\}\equiv\{U\textrm{: there exists }c_{i_{E},k_{E}}^{(E)}c_{j_{E},k_{E}}^{(E)\ast}\neq0,\textrm{ for }i_{E}\neq j_{E}\}$.
If $c_{i_{E},k_{E}}^{(E)}c_{j_{E},k_{E}}^{(E)\ast}\neq0$, Definition
2 implies that $\mathcal{U}$ must transform $|i_{E},E\rangle$ and
$|j_{E},E\rangle$ in superpositions of energy eigenstates (there
also exist $c_{i_{E},l_{E}}^{(E)}\neq0$ and $c_{j_{E},l_{E}}^{(E)}\neq0$,
for $l_{E}\neq k_{E}$). Otherwise, $\langle j_{E},E|\mathcal{U}^{\dagger}\mathcal{U}|i_{E},E\rangle=c_{i_{E},k_{E}}^{(E)}c_{j_{E},k_{E}}^{(E)\ast}\neq0$
and $\mathcal{U}$ would not be unitary. 

\textbf{\textcolor{black}{Theorem~3}}. \textcolor{black}{For $\rho$
arbitrary, the SEC unitary $\tilde{U}$ that maximizes $\Delta_{Diag}\bigl\langle H_{\gamma}\bigr\rangle$
is such that $\Delta_{Coh}\bigl\langle H_{\gamma}\bigr\rangle=0$
\cite{55-generalization of theorems to B}}.

\textit{Role of quantum coherence in the energy-transfer optimization}.\textemdash The
physical impact of Theorem 3 will now become apparent. We start by
pointing out that the coherences in $\rho$ do not contribute to the
initial local energies, namely $\textrm{Tr}\left(H_{\gamma}\chi\right)=\textrm{Tr}_{\gamma}\left(H_{\gamma}\textrm{Tr}_{\gamma'}\chi\right)=0$
for any $\rho$, where $\gamma'=B$ if $\gamma=A$ and viceversa (the
reason for this equality being that $\textrm{Tr}_{\gamma'}\chi$ can
not yield diagonal elements in the eigenbasis of $H_{\gamma}$). Therefore,
the same amount of energy, $\textrm{Tr}\left(H_{\gamma}\rho_{Diag}\right)$,
is initially available in the states $\rho$ and $\rho_{Diag}$ to
be exchanged. Such a property allows us to perform an unbiased comparison
between these states, in order to assess the role that coherence plays
in this task. We find that indeed coherence is a potential resource
to optimize the energy transfer. This is expressed by means of the
inequality 
\begin{equation}
\underset{\{U\}}{\textrm{max}}\Delta\bigl\langle H_{\gamma}\bigr\rangle\geq\underset{\{U\}}{\textrm{max}}\Delta_{Diag}\bigl\langle H_{\gamma}\bigr\rangle,\label{eq:5 (coherence and energy transf)}
\end{equation}
where $\{U\}$ is the full set of SEC unitaries. 

From Eq.~(\ref{eq:4 (decomp of energy transfer)}) and Theorem 3
it follows that, for $U=\tilde{U}$, $\Delta\bigl\langle H_{\gamma}\bigr\rangle=\textrm{max}_{\{U\}}\Delta_{Diag}\bigl\langle H_{\gamma}\bigr\rangle$,
which immediately implies Eq.~(\ref{eq:5 (coherence and energy transf)}).
Theorem 2 provides a necessary condition on the initial coherence,
to obtain an enhancement in the energy transfer (corresponding to
the strict inequality in Eq.\textcolor{red}{~}(\ref{eq:5 (coherence and energy transf)})).
Moreover, Corollary 2.1 tells us that this resource can only be exploited
by some unitary in the set $\{\mathcal{U}\}$\textcolor{black}{. }We
shall\textcolor{black}{{} later corroborate }this quantum thermodynamic
signature\textcolor{black}{{} in the special case of two interacting
qubits. }

\textit{States that only allow energy flow in one direction}.\textemdash Let
us introduce the following set of states: 

\begin{align}
\{\sigma^{(A\leftarrow B)}\} & =\{\rho\textrm{: }\rho_{Diag}^{A}(E)\textrm{ is passive within}\nonumber \\
 & \quad\;\;\;\mathcal{H}_{E}^{A}\textrm{ and }\chi(E,E)=0\textrm{, for all }E\}.\label{eq:6 (unidir energy flow)}
\end{align}
If $\rho\in\{\sigma^{(A\leftarrow B)}\}$, the energy flow for any
SEC unitary occurs from system $B$ to system $A$ (hence the notation
$A\leftarrow B$). The condition of $E$-passivity for any $\rho_{Diag}^{A}(E)$
is necessary and sufficient to have a unidirectional diagonal energy
transfer. In this case, Theorem 1 implies that the energy associated
to each term in the sum for $\Delta_{Diag}\bigl\langle H_{A}\bigr\rangle$
can never decrease. Conversely, if $\rho_{Diag}^{A}(E')$ is not $E'$-passive
on the $E$-local subspace $\mathcal{H}_{E'}^{A}$, we can choose
a set of $E$-local unitaries $\{U_{E}^{A}\}$ such that $U_{E'}^{A}$
reduces the energy of $\rho_{Diag}^{A}(E')$ while the remaining unitaries
are the identity in the corresponding subspaces (for $E\neq E'$).
Theorem 2 guarantees that, for a state $\rho$ containing only coherences
of the form $\chi(E,E')$, with $E\neq E'$, $\Delta\bigl\langle H_{A}\bigr\rangle=\Delta_{Diag}\bigl\langle H_{A}\bigr\rangle$.
Therefore, the total energy transfer to system $A$ is always positive
for the states defined in Eq.\textcolor{red}{~}(\ref{eq:6 (unidir energy flow)}).
On the other hand, we can not assert that this equation encompasses
all the states manifesting unidirectional energy flow. If coherences
$\chi(E,E)\neq0$ are present in $\rho$, answering this question
requires the more involved task of determining the sign of $\Delta\bigl\langle H_{A}\bigr\rangle$. 

The Clausius statement of the second law of thermodynamics applies
to uncorrelated states $\rho_{\beta_{A}}^{A}\otimes\rho_{\beta_{B}}^{B}$,
where $\rho_{\beta_{\gamma}}^{\gamma}$ is a thermal equilibrium state
at inverse temperature $\beta_{\gamma}$ \cite{33-Partovi}. For the
sake of consistency, we prove in \cite{53-Supplementary-Material}
that, for $\beta_{A}>\beta_{B}$, any such state belongs to $\{\sigma^{(A\leftarrow B)}\}$.
However, Eq. (\ref{eq:6 (unidir energy flow)}) evidently extends
the scope of this statement, as it includes states with coherences
of the type $\chi(E,E')$. To further support this generality, we
show in \cite{53-Supplementary-Material} that any tensor product
between a passive state \cite{55-passive,56-pass.states2} of system
$A$ and a maximally active state \cite{57-maxact.state} of system
$B$ also belongs to $\{\sigma^{(A\leftarrow B)}\}$.

\begin{figure}
\centering{}\includegraphics[scale=0.58]{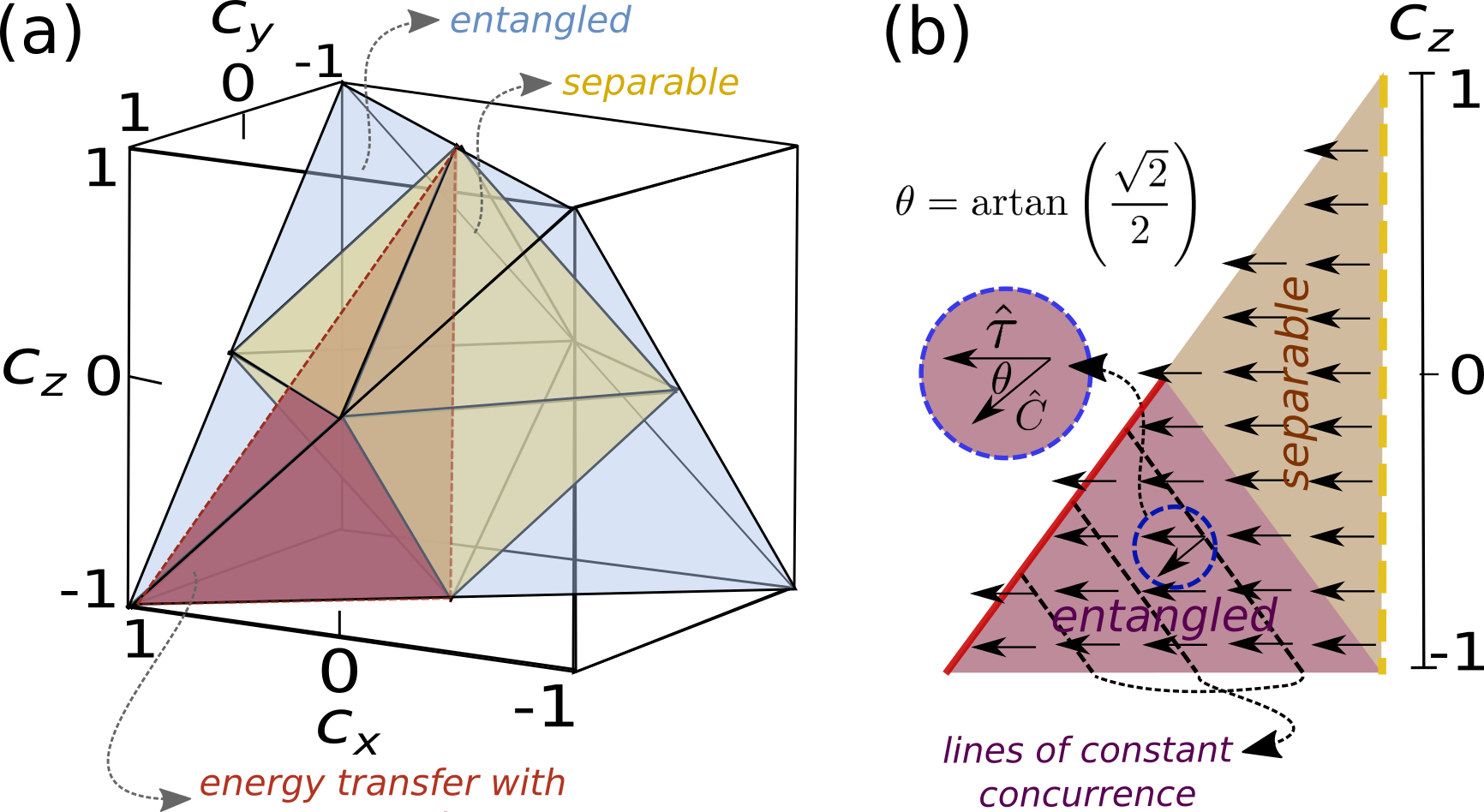}\caption{(a) Energy transfer between two qubits, for initial Bell-diagonal
states. This class can be depicted in terms of the parameters $c_{i}=\textrm{Tr}\left(\sigma_{i}^{A}\otimes\sigma_{i}^{B}\rho\right)$,
where $\left\{ \sigma_{i}^{\gamma}\right\} {}_{i=x,y,z}$ are the
Pauli matrices for the qubit $\gamma$. Separable states lie in the
inner yellow octahedron, while entangled ones lie inside the blue
region and outside the yellow octahedron. The dark red triangle (see
\cite{53-Supplementary-Material}) inside the tetrahedron represents
a subset of states that produce a non null energy transfer, $\Delta\bigl\langle H_{\gamma}\bigr\rangle$,
under a suitable SEC $U$. (b) Frontal view of the aforementioned
subset. On this plane, the maximum energy transfer, $\textrm{max}_{\{U\}}\Delta\bigl\langle H_{\gamma}\bigr\rangle$,
is zero only for classically correlated states (yellow dashed line),
and its gradient has constant projection with direction $\hat{\tau}=1/\sqrt{2}(1,1,0)$.
Within the entangled triangle $\textrm{max}_{\{U\}}\Delta\bigl\langle H_{\gamma}\bigr\rangle$
increases monotonically with the concurrence $C$ (i.e., $\partial\textrm{max}_{\{U\}}\Delta\bigl\langle H_{\gamma}\bigr\rangle/\partial C>0$),
whose projected gradient possesses constant direction $\hat{C}=1/\sqrt{3}(1,1,-1)$
\cite{53-Supplementary-Material}. The left side (continuous red line)
contains maximum coherence states that satisfy Eq.~(\ref{eq:10 (max transfer for bell-diagonal)})
in the main text. }
\end{figure}

\textit{Characterization of the energy exchange between two qubits}.\textemdash{}\textcolor{red}{{}
}We consider here two qubits with identical Hamiltonians $H_{\gamma}=\hbar\omega\left|1\right\rangle {}_{\gamma}\left\langle 1\right|=\left|1\right\rangle {}_{\gamma}\left\langle 1\right|$,
where we set $\hbar\omega=1$ for simplicity and $\left|1\right\rangle {}_{\gamma}$
($\left|0\right\rangle {}_{\gamma}$) represents the excited (ground)
state of qubit $\gamma$. This condition ensures that a SEC unitary
acts non trivially on the energy eigenspace $\mathcal{H}_{E=1}$.
From Theorem~2, only the coherences $\chi(1,1)$ may contribute to
the energy transfer. Therefore, a potential quantum advantage results
from states $\rho=\rho_{Diag}+\chi(1,1),$ where $|\alpha_{0,1}^{(1)}|^{2}\equiv|\alpha_{0,1}^{(1,1)}|^{2}\leq p_{0,1}p_{1,0}$
\cite{52-comment-about-coh} and $p_{i,j}=\textrm{Tr}\left(\left|i\right\rangle {}_{A}\left\langle i\right|\otimes\left|j\right\rangle {}_{B}\left\langle j\right|\rho\right)$
(cf. Eqs.~(\ref{eq:2 (rho_diag)}) and (\ref{eq:3 (rho_coh)})).\textcolor{black}{{}
The description of $\Delta_{Diag}\bigl\langle H_{\gamma}\bigr\rangle$
and $\Delta_{Coh}\bigl\langle H_{\gamma}\bigr\rangle$ is embodied}
by two real parameters, $0\leq r\leq1$ and $0\leq\phi\leq2\pi$,
associated to an arbitrary SEC unitary. If we choose (without loss
of generality) system $A$ \cite{53-Supplementary-Material}, then
\begin{eqnarray}
\Delta_{Diag}\bigl\langle H_{A}\bigr\rangle & = & (p_{0,1}-p_{1,0})r{}^{2}=\left(p_{1}^{B}-p_{1}^{A}\right)r{}^{2},\label{eq:8a (diag transf for two qubits)}\\
\Delta_{Coh}\bigl\langle H_{A}\bigr\rangle & = & 2\textrm{Re}\left(\alpha_{0,1}^{(1)}e^{i\phi}\right)r\sqrt{1-r^{2}},\label{eq:8b (coh transf for two qubits)}
\end{eqnarray}
where ${\color{red}{\normalcolor p_{1}^{\gamma}=}}\textrm{Tr}\bigl(\left|1\right\rangle _{\gamma}\left\langle 1\right|\rho\bigr)$
is the excited population for qubit $\gamma$.\textcolor{red}{{} }

The optimization of $\Delta\bigl\langle H_{\gamma}\bigr\rangle$ yields
\cite{53-Supplementary-Material} 
\begin{equation}
\Delta\bigl\langle H_{\gamma}\bigr\rangle_{\textrm{max}}=\textrm{max}{}_{\{U;\alpha_{0,1}^{(1)}\}}\Delta\bigl\langle H_{\gamma}\bigr\rangle=\mu_{\gamma},\label{eq:9}
\end{equation}
where $\mu_{A}=p_{0,1}$, $\mu_{B}=p_{1,0}$, and the maximum corresponds
to a state of maximum coherence, given by $|\alpha_{0,1}^{(1)}|=\sqrt{p_{0,1}p_{1,0}}$.
In particular, Eq.~(\ref{eq:9}) shows that $\Delta\bigl\langle H_{A}\bigr\rangle_{\textrm{max}}>\textrm{max}_{\{U\}}\Delta_{Diag}\bigl\langle H_{A}\bigr\rangle=p_{0,1}-p_{1,0}$
(for $r=1$ in Eq.~ (\ref{eq:8a (diag transf for two qubits)})).
This represents an enhancement of the energy transfer to system $A$,
due to coherence, and corroborates for two qubits the quantum advantage
suggested by Eq.~(\ref{eq:5 (coherence and energy transf)}). The
exclusive dependence on the local populations $p_{i}^{\gamma}$, expressed
by Eq.~(\ref{eq:8a (diag transf for two qubits)}), also means that
$\Delta_{Diag}\bigl\langle H_{A}\bigr\rangle$ is not affected by
classical correlations in $\rho_{Diag}$. Therefore, for locally thermal
qubits the reversion of the thermodynamic arrow of time is only possible
through the coherent contribution to the energy transfer. If the qubit
$A$ has the larger temperature, such a reversion is implied by the
positive value of $\Delta\bigl\langle H_{A}\bigr\rangle_{\textrm{max}}$
in Eq.~(\ref{eq:9}). We can also interpret this ``anomalous''
heat flow as necessarily owed to quantum correlations. A classically
correlated state $\rho$, with both marginals $\rho^{\gamma}=\textrm{Tr}_{\gamma'}\rho$
being diagonal in $H_{\gamma}$ (as is the case for local thermality),
is a state without coherence in the eigenbasis of $H_{A}+H_{B}$ \cite{58-comment-quantumcorr}.
Hence, the absence of local coherence implies that $\rho$ must have
quantum correlations to produce $\Delta_{Coh}\bigl\langle H_{A}\bigr\rangle\neq0$. 

\textcolor{black}{We complement the discussion about the role of quantum
correlations for energy transfer,} analyzing this process for Bell-diagonal
states \cite{48-belldiag1,49-belldiag2}. The condition of maximally
mixed marginals implies that $p_{0,1}=p_{1,0}$. Therefore, $\Delta\bigl\langle H_{A}\bigr\rangle=\Delta_{Coh}\bigl\langle H_{A}\bigr\rangle$
(cf. Eq.~(\ref{eq:8a (diag transf for two qubits)})) and $\Delta\bigl\langle H_{\gamma}\bigr\rangle_{\textrm{max}}=p_{0,1}$,
according to Eq.\textcolor{red}{~}(\ref{eq:9}). If we specialize
to a subset of entangled states (red line in Fig.\textcolor{red}{~}1
(b)), Eq.\textcolor{red}{~}(\ref{eq:9}) simplifies to \cite{53-Supplementary-Material}
\begin{equation}
\Delta\bigl\langle H_{\gamma}\bigr\rangle_{\textrm{max}}=\frac{1+C(\rho)}{4},\label{eq:10 (max transfer for bell-diagonal)}
\end{equation}
where $C(\rho)=\textrm{max}(0,4p_{0,1}-1)$ is the concurrence. We
note a linear increase of $\Delta\bigl\langle H_{\gamma}\bigr\rangle_{\textrm{max}}$
with respect to $C(\rho)$, which is monotonically associated to the
entanglement of formation \cite{59-Wootters}.\textcolor{black}{{} Figure
1 also depicts separable states that yield $\left|\Delta\bigl\langle H_{\gamma}\bigr\rangle\right|=\left|\Delta_{Coh}\bigl\langle H_{\gamma}\bigr\rangle\right|>0$,
thereby outperforming classically correlated states.}

\textit{Conclusions and perspectives}.\textemdash We have derived
three theorems that constitute a theoretical framework to characterize
the energy transfer in bipartite quantum systems. \textcolor{black}{Theorem
1 describes this process for ``classical'' (incoherent) states,
allowing to obtain the corresponding values for the energy transfer.
Theorem 2 singles out the class of coherence that may have a non null
contribution, as well as the subset of SEC unitaries that could exploit
its potential. Employing Theorem 3 (which follows from Theorems 1
and 2), we showed that the maximum energy transfer (optimized over
the set of SEC unitaries) for a general state is bounded from below
by that of the state dephased in the joint eigenenergy basis. This
implies in particular that, for optimal evolutions, coherence never
worsens the energy exchange. The type of coherence that does have
an impact on this task is also useful for extracting work in a multipartite
scenario, under ``thermal processes'' \cite{60}. Further investigations
on such connection are pertinent.}\textcolor{red}{{} }On the other hand,
we employed Theorems 1 and 2 to deduce a novel class of states that
only allow \textcolor{black}{unidirectional energy flow}. An open
question is whether this set includes all bipartite states satisfying
the mentioned constraint. 

We illustrated our results describing the energy transfer between
two qubits. In this case, coherence provides a genuine quantum advantage
over incoherent states. It is also the fundamental resource for reverting
the thermodynamic arrow of time, in connection with the experimental
\textcolor{black}{findings} reported in \cite{35-exp-rev-of-arrow}.
\textcolor{black}{For Bell diagonal states, we found that entanglement
is not necessary to outperform classically correlated states. Moreover,
both entangled and separable states provide a quantum enhancement
only if the state contains ``useful'' coherence,}\textcolor{red}{{}
}characterized by Theorem 2\textcolor{black}{. For a suitable subset
of entangled states, the maximum energy transfer increases monotonically
with the concurrence. In particular, Eq.}\textcolor{red}{~}\textcolor{black}{(\ref{eq:10 (max transfer for bell-diagonal)})
exemplifies this behavior for entangled states of maximum coherence.
Searching for a similar relation in systems of higher dimension could
be an interesting extension to this analysis.}
\begin{acknowledgments}
We thank P. Camati for very fruitful discussions. We acknowledge financial
support from UFABC, CNPq, CAPES, and FAPESP. R.M.S. gratefully acknowledges
financial support from the Royal Society through the Newton Advanced
Fellowship scheme (Grant no.~NA140436). This research was performed
as part of the Brazilian National Institute of Science and Technology
for Quantum Information (INCT-IQ). 
\end{acknowledgments}

\section*{Supplementary Material}

\global\long\def\thesection{S-\Roman{section}}
\setcounter{section}{0}
\global\long\def\thefigure{S\arabic{figure}}
\setcounter{figure}{0}
\global\long\def\theequation{S\arabic{equation}}
\setcounter{equation}{0}

\subsection*{Proofs of Theorems 1-3}

\textbf{\textit{Proof of Theorem 1}}: Replacing Eq.~(2) of the main
text in the expression for $\Delta_{Diag}\bigl\langle H_{A}\bigr\rangle$
(cf. Eq.~(4) of the main text) we obtain the equation $\Delta_{Diag}\bigl\langle H_{A}\bigr\rangle=\sum_{E}p_{E}\textrm{Tr}_{A}\left(H_{A}\textrm{Tr}_{B}(U\rho_{Diag}(E)U^{\dagger})-H_{A}\rho_{Diag}^{A}(E)\right)$,
where $\rho_{Diag}^{A}(E)=\textrm{Tr}_{B}\rho_{Diag}(E)$. From Definition
2 and Eq.~(2) of the main text, $U\rho_{Diag}(E)U^{\dagger}=\sum_{i_{E};j_{E},k_{E}}c_{i_{E},j_{E}}^{(E)}c_{i_{E},k_{E}}^{(E)\ast}\frac{p(\varepsilon_{i_{E}},E)}{p_{E}}\Pi_{j_{E},k_{E}}^{(E)},$
where $\Pi_{j_{E},k_{E}}^{(E)}=\Pi_{j_{E},k_{E}}^{(E,E)}$ is defined
through Eq.~(3) of the main text. Using Definition 2 we find a similar
expression for $U_{E}^{A}\rho_{Diag}^{A}(E)U_{E}^{A\dagger}$: $U_{E}^{A}\rho_{Diag}^{A}(E)U_{E}^{A\dagger}=\sum_{i_{E};j_{E},k_{E}}c_{i_{E},j_{E}}^{(E)}c_{i_{E},k_{E}}^{(E)\ast}\frac{p(\varepsilon_{i_{E}},E)}{p_{E}}|j_{E}\rangle_{A}\langle k_{E}|.$
\\

The expressions for $U\rho_{Diag}(E)U^{\dagger}$ and $U_{E}^{A}\rho_{Diag}^{A}(E)U_{E}^{A\dagger}$
are further related by the identity $\textrm{Tr}_{B}\left(U\rho_{Diag}(E)U^{\dagger}\right)=\mathcal{D}_{H_{A}}\left(U_{E}^{A}\rho_{Diag}^{A}(E)U_{E}^{A\dagger}\right),$
where $\mathcal{D}_{H_{A}}(\cdot)$ is the map that eliminates all
coherences in the eigenbasis of $H_{A}$, while leaving unmodified
the populations (dephasing with respect to $H_{A}$). To derive this
equality we must compute the operators $\textrm{Tr}_{B}\Pi_{j_{E},k_{E}}^{(E)}$,
appearing in $\textrm{Tr}_{B}\left(U\rho_{Diag}(E)U^{\dagger}\right)$.
Instead of using the notation of the main text, $\Pi_{j_{E},k_{E}}^{(E)}=|j_{E},E\rangle_{A}\langle k_{E},E|$,
it is convenient to write $\Pi_{j_{E},k_{E}}^{(E)}$ as $\Pi_{j_{E},k_{E}}^{(E)}=|j_{E}\rangle_{A}\langle k_{E}|\otimes|j'_{E}\rangle_{B}\langle k'_{E}|$
(recall that for any $j_{E}$ $(k_{E})$ the corresponding $j'_{E}$
($k'_{E}$) has a unique value due to the non-degeneracy of $H_{B}$).
In this way, $\textrm{Tr}_{B}\Pi_{j_{E},k_{E}}^{(E)}=|j_{E}\rangle_{A}\langle k_{E}|\delta_{j'_{E},k'_{E}}$.
Now we show that, \textit{under the constraint of non-degeneracy for
$H_{A}$ and $H_{B}$}, $\delta_{j'_{E},k'_{E}}=\delta_{j_{E},k{}_{E}}$.
By definition, the eigenenergies of $|j_{E}\rangle_{A}$ and $|k_{E}\rangle_{A}$
are related to those of $|j'_{E}\rangle_{A}$ and $|k'_{E}\rangle_{A}$
through the equations $\varepsilon_{j_{E}}=E-\bar{\varepsilon}_{j'_{E}}$
and $\varepsilon_{k_{E}}=E-\bar{\varepsilon}_{k'_{E}}$. Hence, if
$j'_{E}=k'_{E}$, the non-degeneracy of $H_{A}$ implies that $j_{E}=k{}_{E}$.
Conversely, for $j_{E}=k{}_{E}$, the non-degeneracy of $H_{B}$ implies
that $j'_{E}=k'{}_{E}$. Therefore, $\delta_{j'_{E},k'_{E}}=\delta_{j_{E},k{}_{E}}$
and $\textrm{Tr}_{B}\Pi_{j_{E},k_{E}}^{(E)}=|j_{E}\rangle_{A}\langle k_{E}|\delta_{j_{E},k_{E}}$,
which, after substitution in $\textrm{Tr}_{B}\left(U\rho_{Diag}(E)U^{\dagger}\right)$,
yields: 
\begin{align*}
\textrm{Tr}_{B}\left(U\rho_{Diag}(E)U^{\dagger}\right) & =\sum_{i_{E},j_{E}}|c_{i_{E},j_{E}}^{(E)}|^{2}\frac{p(\varepsilon_{i_{E}},E)}{p_{E}}|j_{E}\rangle_{A}\langle j_{E}|\\
 & =\mathcal{D}_{H_{A}}\left(U_{E}^{A}\rho_{Diag}^{A}(E)U_{E}^{A\dagger}\right).
\end{align*}
The proof is concluded by noticing that $\mathcal{D}_{H_{A}}(\cdot)$
does not modify the average energy of system $A$. \\

\textbf{\textit{Proof of Theorem }}\textbf{2}: According to Eq.~(4)
of the main text, $\Delta_{Coh}\bigl\langle H_{A}\bigr\rangle={\color{red}{\normalcolor \textrm{Tr}\left(H_{A}U\chi U^{\dagger}\right)}}-\textrm{Tr}\left(H_{A}\chi\right)$.
Employing again the notation $\Pi_{i_{E},j_{E'}}^{(E,E')}=|i_{E}\rangle_{A}\langle j_{E'}|\otimes|k{}_{E}\rangle_{B}\langle l{}_{E'}|$,
used in the previous proof, it is easily shown that $\textrm{Tr}\left(H_{A}\chi\right)=0$:
For any $\Pi_{i_{E},j_{E'}}^{(E,E')}$ we have that $\textrm{Tr}_{A}\left(H_{A}\textrm{Tr}_{B}\Pi_{i_{E},j_{E'}}^{(E,E')}\right)=\textrm{Tr}_{A}\left(H_{A}|i_{E}\rangle_{A}\langle j_{E'}|\delta_{k_{E},l_{E'}}\right).$
This expression equals zero if either $i_{E}\neq j_{E'}$ or $k_{E}\neq l_{E'}$.
Therefore, $\textrm{Tr}\left(H_{A}\chi\right)=0$.\\

On the other hand, 
\[
U\Pi_{i_{E},j_{E'}}^{(E,E')}U^{\dagger}=\sum_{k_{E}.l_{E'}}c_{i_{E},k_{E}}^{(E)}c_{j_{E'},l_{E'}}^{(E')\ast}\Pi_{k_{E},l_{E'}}^{(E,E')},
\]
applying Definition 2. Since for $E\neq E'$ all the $\Pi_{k_{E},l_{E'}}^{(E,E')}$
are global coherent elements, $\sum_{E,E':\,E\neq E'}\textrm{Tr}\left(H_{A}U\chi(E,E')U^{\dagger}\right)=0$.
For coherences of the type $\chi(E,E)=\sum_{i_{E},j_{E}:\,i_{E}\neq j_{E}}{\color{red}{\normalcolor \alpha_{i_{E},j_{E}}^{(E,E)}}}\Pi_{i_{E},j_{E}}^{(E,E)}$
(cf. Eq.~(3) of the main text) we obtain
\[
\textrm{Tr}_{B}\sum_{E}U\chi(E,E)U^{\dagger}=\sum_{E}\sum_{i_{E}\neq j_{E}}{\color{red}{\normalcolor \alpha_{i_{E},j_{E}}^{(E,E)}}}\textrm{Tr}_{B}U\Pi_{i_{E},j_{E}}^{(E,E)}U^{\dagger}.
\]
From~the~relation~$\textrm{Tr}_{B}\Pi_{k_{E},l_{E}}^{(E,E)}=|k_{E}\rangle_{A}\langle l_{E}|\delta_{k_{E},l_{E}}$,
derived in the previous proof, 
\[
\textrm{Tr}_{B}U\Pi_{i_{E},j_{E}}^{(E,E)}U^{\dagger}=\sum_{k_{E}}c_{i_{E},k_{E}}^{(E)}c_{j_{E},k_{E}}^{(E)\ast}|k_{E}\rangle_{A}\langle k_{E}|.
\]
Therefore, $\textrm{Tr}_{B}\sum_{E}U\chi(E,E)U^{\dagger}$ equals
$\sum_{E}\sum_{k_{E}}\sum_{i_{E}<j_{E}}2\textrm{Re}\left({\color{red}{\normalcolor \alpha_{i_{E},j_{E}}^{(E,E)}}}c_{i_{E},k_{E}}^{(E)}c_{j_{E},k_{E}}^{(E)\ast}\right)|k_{E}\rangle_{A}\langle k_{E}|$,
after inverting the order of the sums $\sum_{i_{E}\neq j_{E}}$ and
$\sum_{k_{E}}$. In this expression, the index $E$ runs freely over
the eigenvalues of $H$ and the sum $\sum_{k_{E}}$ runs over values
of $k_{E}$ such that the state $|k_{E},E\rangle$ (with eigenenergies
$\varepsilon_{k_{E}}$ and $E$) exists. We can also invert the order
for the sums $\sum_{E}$ and $\sum_{k_{E}}$ keeping in mind this
constraint. The resulting expression is $\textrm{Tr}_{B}\sum_{E}U\chi(E,E)U^{\dagger}=\sum_{k}\eta_{k}|k\rangle_{A}\langle k|$,
where $\eta_{k}\equiv\sum'_{E}\sum_{i_{E}<j_{E}}2\textrm{Re}\left({\color{red}{\normalcolor \alpha_{i_{E},j_{E}}^{(E,E)}}}c_{i_{E},k}^{(E)}c_{j_{E},k}^{(E)\ast}\right)$.
Now the index $k$ runs freely over the eigenvalues of $H_{A}$ and
the aforementioned constraint results from restricting the sum over
$E$: for any $k$, $\sum'_{E}$ is restricted to values of $E$ such
that the state $|k,E\rangle$ (with eigenenergies $\varepsilon_{k}$
and $E$) exists. In this way we ensure that the sums $\sum_{E}\sum_{k_{E}}$
and $\sum_{k}\sum'_{E}$ cover exactly the same terms. Therefore,
$\Delta_{Coh}\bigl\langle H_{A}\bigr\rangle=\textrm{Tr}_{A}H_{A}\left(\textrm{Tr}_{B}\sum_{E}U\chi(E,E)U^{\dagger}\right)=\sum_{k}\eta_{k}\varepsilon_{k}$.
\\

\textbf{\textit{Proof of Theorem }}\textbf{3}: From Theorem 1 and
Definitions 1 and 2 of the main text, $\Delta_{Diag}\bigl\langle H_{A}\bigr\rangle$
can be maximized by independently maximizing each term $\textrm{Tr}_{A}\left(H_{A}U_{E}^{A}\rho_{Diag}^{A}(E)U_{E}^{A\dagger}\right)$,
with respect to $U_{E}^{A}$. The solution corresponds to $\tilde{U}_{E}^{A}$
such that $\tilde{U}_{E}^{A}\rho_{Diag}^{A}(E)\tilde{U}_{E}^{A\dagger}$
is the $E$-local state of maximum energy, obtained from $\rho_{Diag}^{A}(E)$
through an $E$-local unitary (Definition 3). Since both $\rho_{Diag}^{A}(E)=\textrm{Tr}_{B}\rho_{Diag}(E)$
and $\tilde{U}_{E}^{A}\rho_{Diag}^{A}(E)\tilde{U}_{E}^{A\dagger}$
are diagonal in the eigenbasis of $H_{A}$, $\tilde{U}_{E}^{A}\left|i_{E}\right\rangle _{A}=\left|\tilde{k}_{E}\right\rangle _{A}$.
The corresponding coefficients $\tilde{c}_{i_{E},k_{E}}^{(E)}$ (see
Definition 2) satisfy the simple relation $\tilde{c}_{i_{E},k_{E}}^{(E)}=\delta_{k_{E},\tilde{k}_{E}}$.
This implies that $\tilde{c}_{i_{E},k_{E}}^{(E)}\tilde{c}_{i'_{E},k_{E}}^{(E)\ast}=\delta_{k_{E},\tilde{k}_{E}}\delta_{k_{E},\tilde{k}'_{E}}=\delta_{\tilde{k}_{E},\tilde{k}'_{E}}=0$
for any pair $(\tilde{c}_{i_{E},k_{E}}^{(E)},\tilde{c}_{i'_{E},k_{E}}^{(E)})$,
given the unitary character of $\tilde{U}_{E}^{A}$ (otherwise, $\tilde{U}_{E}^{A}\left|i_{E}\right\rangle _{A}=\tilde{U}_{E}^{A}\left|i'_{E}\right\rangle _{A}=\left|\tilde{k}_{E}\right\rangle _{A}$,
resulting in a non unitary map). Therefore, $\tilde{U}\notin\{\mathcal{U}\}$
and according to Corollary 2.1 $\Delta_{Coh}\bigl\langle H_{A}\bigr\rangle=0$. 

\subsection*{\textit{\emph{Direction of energy flow for tensor products between
thermal states and between a passive state and a maximally active
one}}}

Let us consider a tensor product of the form $\rho=\rho_{\beta_{A}}^{A}\otimes\rho_{\beta_{B}}^{B}$,
with 
\begin{equation}
\rho_{\beta_{A}}^{A}=\frac{\textrm{exp}(-\beta_{A}H_{A})}{Z_{\beta_{A}}^{A}},\quad\rho_{\beta_{B}}^{B}=\frac{\textrm{exp}(-\beta_{B}H_{B})}{Z_{\beta_{B}}^{B}},\label{eq:1}
\end{equation}
thermal equilibrium states at inverse temperatures $\beta_{A}$ and
$\beta_{B}$, respectively; where $Z_{\beta_{\gamma}}^{\gamma}=\textrm{Tr}\left[\textrm{exp}(-\beta_{\gamma}H_{\gamma})\right]$
is the partition function. Owing to the non-degeneracy condition of
the local hamiltonians ($H_{A}$ and $H_{B}$), the eigenvalues of
the joint state restricted to the eigenspace of energy $E$ ($\mathcal{H}^{E}$),
$\rho_{Diag}(E)$, and the eigenvalues of the corresponding $E$-local
state, $\rho_{Diag}^{A}(E)=\textrm{Tr}_{B}\rho_{Diag}(E)$, are identical
(cf. Eq.~(3) of the main text). These eigenvalues are explicitly
given by 
\begin{equation}
\lambda_{i_{E}}=\frac{\textrm{exp}(-\beta_{B}E)}{p_{E}}\frac{\textrm{exp}[-(\beta_{A}-\beta_{B})\varepsilon_{i_{E}}]}{{\normalcolor {\normalcolor {\color{red}{\normalcolor Z_{\beta_{A}}^{A}}{\normalcolor Z}_{{\normalcolor \beta_{B}}}^{{\normalcolor B}}}}}}.\label{eq:2}
\end{equation}

If the system $A$ has lower temperature than system $B$, then $\beta_{A}>\beta_{B}$
and $\lambda_{i_{E}}<\lambda_{i'_{E}}$ for $\varepsilon_{i_{E}}>\varepsilon_{i'_{E}}$.
Therefore, for any value of $E$ the state $\rho_{Diag}^{A}(E)$ is
passive within $\mathcal{H}_{E}^{A}$. Since $\rho_{\beta_{A}}^{A}\otimes\rho_{\beta_{B}}^{B}$
is diagonal, we conclude from Theorem~1 and Eq.~ (6) of the main
text that $\Delta\bigl\langle H_{A}\bigr\rangle=\Delta_{Diag}\bigl\langle H_{A}\bigr\rangle>0$
for any SEC unitary, meaning that heat can only flow from the hotter
system ($B$) to the colder one ($A$).

On the other hand, consider now the product $\rho=\rho_{P}^{A}\otimes\rho_{M}^{B}$,
where 

\begin{equation}
\rho_{P}^{A}=\sum_{i}\lambda_{i}\left|\varepsilon_{i}\right\rangle {}_{A}\left\langle \varepsilon_{i}\right|\label{eq:3}
\end{equation}
is a passive ($P$) state for system $A$ and 
\begin{equation}
\rho_{M}^{B}=\sum_{j}\eta_{j}\left|\bar{\varepsilon}_{j}\right\rangle {}_{B}\left\langle \bar{\varepsilon}_{j}\right|\label{eq:4}
\end{equation}
is a maximally active ($M$) state for system $B$. This formally
means that $\lambda_{i+1}\leq\lambda_{i}$ and $\eta_{j+1}\geq\eta_{j}$
for all $i,j$, for eigenenergies put in increasing order: $\varepsilon_{i+1}\geq\varepsilon_{i}$
and $\varepsilon_{j+1}\geq\varepsilon_{j}$. We immediately note that
$\rho=\rho_{Diag}$ and therefore $\Delta\bigl\langle H_{A}\bigr\rangle=\Delta_{Diag}\bigl\langle H_{A}\bigr\rangle$. 

Keeping in mind the constraint of non-degeneracy of the local hamiltonians,
we can express the state $\rho_{Diag}(E)$ as 
\begin{equation}
\rho_{Diag}(E)=\sum_{i_{E}}\frac{\lambda_{i_{E}}\eta'_{i_{E}}}{p_{E}}\left|\varepsilon_{i_{E}}\right\rangle _{A}\left\langle \varepsilon_{i_{E}}\right|\otimes\left|E-\varepsilon_{i_{E}}\right\rangle _{B}\left\langle E-\varepsilon_{i_{E}}\right|,\label{eq:5}
\end{equation}
where $\eta'_{i_{E}}$ is the eigenvalue of $\rho_{M}^{B}$ corresponding
to the eigenstate $\left|E-\varepsilon_{i_{E}}\right\rangle _{B}$,
with eigenergy $E-\varepsilon_{i_{E}}$, and $p_{E}=\sum_{i_{E}}\lambda_{i_{E}}\eta'_{i_{E}}$.
The property of passivity implies in particular that $\lambda_{i_{E}+1}\leq\lambda_{i_{E}}$.
Likewise, $E-\varepsilon_{i_{E}}>E-\varepsilon_{i_{E}+1}$, by definition,
and therefore $\tilde{\eta}_{i_{E}}\geq\tilde{\eta}_{i_{E}+1}$, given
that $\rho_{M}^{B}$ is maximally active. In this way, the eigenvalues
of $\rho_{Diag}(E)$, $\{\lambda_{i_{E}}\eta'_{i_{E}}/p_{E}\}$, are
monotonically decreasing. This implies that for any $E$ the state
$\rho_{Diag}^{A}(E)=\textrm{Tr}_{B}\rho_{Diag}(E)=\frac{1}{p_{E}}\sum_{i_{E}}\lambda_{i_{E}}\eta'_{i_{E}}\left|\varepsilon_{i_{E}}\right\rangle _{A}\left\langle \varepsilon_{i_{E}}\right|$
is $E$-passive (cf. Definition 3). From Theorem 1 and Eq.~ (6) of
the main text it follows that for this class of states energy can
only be transferred from system $A$ to system $B$. 

\subsection*{Energy transfer for two-qubit states}

A general SEC unitary acting on two-qubit states can be parametrized
as the non-trivial transformation

\begin{align}
U\left|0,0\right\rangle  & =e^{i\theta_{0,0}}\left|0,0\right\rangle ,\label{eq:S6}\\
U\left|1,1\right\rangle  & =e^{i\theta_{1,1}}\left|1,1\right\rangle ,\\
U\left|0,1\right\rangle  & =e^{i\theta_{0,1}}\left(\sqrt{1-r^{2}}\left|0,1\right\rangle +re^{i\varphi}\left|1,0\right\rangle \right),\\
U\left|1,0\right\rangle  & =e^{i\theta_{1,0}}\left(\sqrt{1-r^{2}}\left|1,0\right\rangle -re^{-i\varphi}\left|0,1\right\rangle \right),\label{eq:S9}
\end{align}
where $0\leq r\leq1$ and $\{\theta_{i,j},\varphi\}_{0\leq i,j\leq1}$
are phases in the interval $(0,2\pi)$. Notice that besides fulfilling
Eqs.~(\ref{eq:S6})-(\ref{eq:S9}), the energy gaps of both qubits
must coincide for $U$ to be SEC. 

By applying Eqs.~(\ref{eq:S6})-(\ref{eq:S9}) to the diagonal part
of a two-qubit state, a bit of algebra leads to the following expression
for the transformed local state $\textrm{Tr}_{B}\left(U\rho_{Diag}U^{\dagger}\right)$:
\begin{align}
\textrm{Tr}_{B}\left(U\rho_{Diag}U^{\dagger}\right) & =\left(p_{0,0}+p_{0,1}(1-r^{2})+p_{1,0}r^{2}\right)\left|0\right\rangle {}_{A}\left\langle 0\right|\nonumber \\
 & +[p_{1,1}+p_{0,1}r^{2}+p_{1,0}(1-r^{2})]\left|1\right\rangle {}_{A}\left\langle 1\right|,\label{eq:7}
\end{align}
where $p_{i,j}=\textrm{Tr}\left(\left|i\right\rangle {}_{A}\left\langle i\right|\otimes\left|j\right\rangle {}_{B}\left\langle j\right|\rho_{Diag}\right)$.
On the other hand, $\textrm{Tr}_{B}\left(\rho_{Diag}\right)=(p_{0,0}+p_{0,1})\left|0\right\rangle {}_{A}\left\langle 0\right|+(p_{1,1}+p_{1,0})\left|1\right\rangle {}_{A}\left\langle 1\right|.$
Therefore, using the definition of $\Delta_{Diag}\bigl\langle H_{A}\bigr\rangle$,
$\Delta_{Diag}\bigl\langle H_{A}\bigr\rangle\equiv\textrm{Tr}\left(H_{A}U\rho_{Diag}U^{\dagger}\right)-\textrm{Tr}\left(H_{A}\rho_{Diag}\right)$,
we get 
\begin{equation}
\Delta_{Diag}\bigl\langle H_{A}\bigr\rangle=(p_{0,1}-p_{1,0})r^{2}\hbar\omega=(p_{1}^{B}-p_{1}^{A})r^{2}\hbar\omega,\label{eq:8}
\end{equation}
where $p_{1}^{\gamma}=\textrm{Tr}\left(|1\rangle_{\gamma}\langle1|\rho_{Diag}\right)$
is the population of the excited state for qubit $\gamma$. 

The energy contribution from the ``useful'' coherences of the joint
state, $\chi(\hbar\omega,\hbar\omega)$, is obtained by means of the
transformations $U(|0\rangle_{A}\langle1|\otimes|1\rangle_{B}\langle0|)U^{\dagger}$
and $U(|1\rangle_{A}\langle0|\otimes|0\rangle_{B}\langle1|)U^{\dagger}$.
Employing again Eqs.~(\ref{eq:S6})-(\ref{eq:S9}) we find that 

\begin{equation}
\textrm{Tr}_{B}\left(U\chi(\hbar\omega,\hbar\omega)U^{\dagger}\right)=2\textrm{Re}(\alpha_{0,1}^{(1)}e^{i\phi})r\sqrt{1-r^{2}}\sigma_{Z}^{A},\label{eq:9-1}
\end{equation}
where $\phi\equiv\theta_{0,1}-\theta_{1,0}+\varphi$. In this way, 

\begin{equation}
\begin{array}{ccc}
\Delta_{Coh}\bigl\langle H_{A}\bigr\rangle & = & \textrm{Tr}\left(H_{A}U\chi(\hbar\omega,\hbar\omega)U^{\dagger}\right)\\
 & = & 2\textrm{Re}(\alpha_{0,1}^{(1)}e^{i\phi})r\sqrt{1-r^{2}}\hbar\omega.
\end{array}\label{eq:10}
\end{equation}

\subsection*{Illustrative example: Maximum energy transfer between two qubits}

To obtain Eq.~(9) of the main text, we first maximize $\Delta\bigl\langle H_{\gamma}\bigr\rangle$
with respect to $\alpha_{0,1}^{(1)}$ and $\phi$, for $r$ fixed.
From the condition of energy conservation, $\Delta\bigl\langle H_{A}\bigr\rangle=-\Delta\bigl\langle H_{B}\bigr\rangle$,
and we can maximize $\Delta\bigl\langle H_{B}\bigr\rangle$ by minimizing
$\Delta\bigl\langle H_{A}\bigr\rangle$. The optimization with respect
to $\alpha_{0,1}^{(1)}$ and $\phi$ only encompasses the coherent
energy transfer. For $r$ fixed, the maximun and minimun of $\Delta_{Coh}\bigl\langle H_{A}\bigr\rangle$
are given by ${\normalcolor 2r\sqrt{p_{0,1}p_{1,0}\left(1-r^{2}\right)}}\hbar\omega$
and $-2r\sqrt{p_{0,1}p_{1,0}\left(1-r^{2}\right)}\hbar\omega$, respectively.
These values are obtained from Eq. (\ref{eq:10}), by choosing $\phi=0$
and $\alpha_{0,1}^{(1)}=\pm\sqrt{p_{0,1}p_{1,0}}$. Therefore, $\textrm{max}_{\{r,\phi;\alpha_{0,1}^{(1)}\}}\Delta\bigl\langle H_{A}\bigr\rangle=\textrm{max}_{r}\Delta\bigl\langle H_{A}\bigr\rangle_{+}$,
where ${\normalcolor \Delta\bigl\langle H_{A}\bigr\rangle_{+}/\hbar\omega\equiv(p_{0,1}-p_{1,0})r^{2}+}{\normalcolor 2r\sqrt{p_{0,1}p_{1,0}\left(1-r^{2}\right)}}$.
Similarly, $\textrm{max}_{\{r,\phi;\alpha_{0,1}^{(1)}\}}\Delta\bigl\langle H_{B}\bigr\rangle=\textrm{min}_{r}\Delta\bigl\langle H_{A}\bigr\rangle_{-}$,
with ${\normalcolor \Delta\bigl\langle H_{A}\bigr\rangle_{-}/\hbar\omega\equiv(p_{0,1}-p_{1,0})r^{2}-}{\normalcolor 2r\sqrt{p_{0,1}p_{1,0}\left(1-r^{2}\right)}}$. 

By employing the chain rule, we find that $\frac{\partial}{\partial r}\Delta\bigl\langle H_{A}\bigr\rangle_{\pm}=2r\frac{\partial}{\partial x}{\color{red}{\normalcolor \Delta\bigl\langle H_{A}\bigr\rangle_{\pm}}}$,
being $x\equiv r^{2}$. Since for $r=0$ we get $\Delta\bigl\langle H_{A}\bigr\rangle_{\pm}=0$,
according to Eqs.~(\ref{eq:8}) and (\ref{eq:10}), the values of
$r$ that yield the optimization are the solutions of the equation
\begin{align}
\frac{1}{\hbar\omega}\frac{\partial}{\partial x}\Delta\bigl\langle H_{A}\bigr\rangle_{\pm} & =(p_{0,1}-p_{1,0})\nonumber \\
 & \pm\frac{\sqrt{p_{0,1}p_{1,0}}}{\sqrt{x(1-x)}}(1-2x)=0.\label{eq:11}
\end{align}
This expression can be rewritten as 

\begin{equation}
(p_{0,1}+p_{1,0})^{2}x^{2}-(p_{0,1}+p_{1,0})^{2}x+p_{0,1}p_{1,0}=0.\label{eq:12}
\end{equation}
The solutions of Eq.~(\ref{eq:12}) are:
\begin{align}
x_{+} & =\frac{p_{0,1}}{p_{0,1}+p_{1,0}},\label{eq:13}\\
x_{-} & =\frac{p_{1,0}}{p_{0,1}+p_{1,0}},\label{eq:13.1}
\end{align}
where $x_{+}$ satisfies $\frac{\partial}{\partial x}\Delta\bigl\langle H_{A}\bigr\rangle_{+}=0$
and $\frac{\partial}{\partial x}\Delta\bigl\langle H_{A}\bigr\rangle_{-}=0$
for $x_{-}$.

Now we verify that $\Delta\bigl\langle H_{A}\bigr\rangle_{+}(x_{+})$
is a maximum and that $\Delta\bigl\langle H_{A}\bigr\rangle_{-}(x_{-})$
corresponds to a minimum. The second derivative with respect to $r$
yields 

\begin{equation}
\frac{\partial^{2}}{\partial r^{2}}\Delta\bigl\langle H_{A}\bigr\rangle_{\pm}=2\frac{\partial\Delta\bigl\langle H_{A}\bigr\rangle_{\pm}}{\partial x}+4x\frac{\partial^{2}\Delta\bigl\langle H_{A}\bigr\rangle_{\pm}}{\partial x^{2}},\label{eq:14}
\end{equation}
where $\frac{1}{\hbar\omega}\frac{\partial^{2}\Delta\bigl\langle H_{A}\bigr\rangle_{\pm}}{\partial x^{2}}=\mp\frac{1}{2}\frac{\sqrt{p_{0,1}p_{1,0}}}{[x(1-x)]^{3/2}}$.
Therefore, $\frac{\partial^{2}\Delta\bigl\langle H_{A}\bigr\rangle_{+}}{\partial r^{2}}\Bigl|_{x_{+}}<0$
and $\frac{\partial^{2}\Delta\bigl\langle H_{A}\bigr\rangle_{-}}{\partial x^{2}}\Bigl|_{x_{-}}>0$
, as expected. By replacing the expressions of Eqs.~(\ref{eq:13})
and (\ref{eq:13.1}) into $\Delta\bigl\langle H_{A}\bigr\rangle_{+}(x_{+})$
and $\Delta\bigl\langle H_{A}\bigr\rangle_{-}(x_{-})$, we arrive
at Eq.~(9) of the main text. 

\subsection*{Energy transfer for Bell-diagonal states}

We analyse here Bell-diagonal states that satisfy the equation $\rho=\rho_{Diag}+\chi(\hbar\omega,\hbar\omega)$,
since for these states $\Delta_{Coh}\bigl\langle H_{A}\bigr\rangle>0$
(cf. Eq.~ (\ref{eq:10})). For the considered class the marginals
of both qubits are maximally mixed states. If we denote as $p_{i}^{\gamma}$,
$i=0,1$, the local populations for qubit $\gamma$, then $p_{0}^{\gamma}=p_{1}^{\gamma}=1/2$.
Therefore, the diagonal energy transfer is equal to zero, according
to Eq.~(\ref{eq:8}), and any value $\Delta\bigl\langle H_{A}\bigr\rangle\neq0$
is associated to a quantum advantage. On the other hand, 
\begin{align*}
p_{0}^{A} & =p_{0,0}+p_{0,1},\\
p_{1}^{A} & =p_{1,1}+p_{1,0},\\
p_{0}^{B} & =p_{0,0}+p_{1,0},\\
p_{1}^{B} & =p_{1,1}+p_{0,1},
\end{align*}
where $p_{i,j}=\textrm{Tr}\left(\left|i\right\rangle {}_{A}\left\langle i\right|\otimes\left|j\right\rangle {}_{B}\left\langle j\right|\rho\right)$.
In this way, the condition of maximally mixed marginals implies that
\begin{equation}
p_{0,1}=p_{1,0},\quad p_{0,0}=p_{1,1}.\label{eq:16}
\end{equation}

Bell diagonal states have a Bloch representation through the expression
$\rho=\frac{1}{4}\left(\mathbb{I}+\sum_{i=x,y,z}c_{i}\sigma_{i}^{A}\otimes\sigma_{i}^{B}\right)$,
where $c_{i}=\textrm{Tr}\left(\sigma_{i}^{A}\otimes\sigma_{i}^{B}\rho\right)$
and $\left\{ \sigma_{i}^{\gamma}\right\} {}_{i=x,y,z}$ are the Pauli
matrices for the qubit $\gamma$. For the states of interest we obtain: 

\begin{align}
c_{x} & =c_{y}=\textrm{Tr}\left(\sigma_{x}^{A}\otimes\sigma_{x}^{B}\chi\right)=2\textrm{Re}(\alpha_{0,1}^{(1)}),\label{eq:19}\\
c_{z} & =\textrm{Tr}\left(\sigma_{z}^{A}\otimes\sigma_{z}^{B}\rho_{Diag}\right)=1-4p_{0,1}.\label{eq:20}
\end{align}
In addition, to demand that such states be Bell diagonal, we require
that $\textrm{Tr}\left(\sigma_{i}^{A}\otimes\sigma_{j}^{B}\chi\right)=0$,
for $i\neq j$. By performing a direct computation we verify that
this imposes the condition $\textrm{Im}(\alpha_{0,1}^{(1)})=0$.

Figure 1 (a) of the main text shows a graphical depiction of Bell
diagonal states, which is constructed by means of a cartesian coordinate
system with axes $c_{i}$. In this parameter space, Eqs. ~(\ref{eq:19})
and ~(\ref{eq:20}) describe a plane that intersects the tetrahedron
and passes through the points $(c_{x},c_{y},c_{z})=(0,0,1)$, $(1,1,-1)$
and $(-1,-1,-1)$. This whole region contains all the two-qubit states
that satisfy the relation $\rho=\rho_{Diag}+\chi(\hbar\omega,\hbar\omega)$,
while simultaneously being Bell-diagonal. On the other hand, it is
sufficient to focus on half of such a region to describe the associated
energy transfer, as we explain next. First, notice that if we choose
$\alpha_{0,1}^{(1)}$ not only real but also positive, it is still
possible to obtain any value of $\Delta\bigl\langle H_{A}\bigr\rangle=\Delta_{Coh}\bigl\langle H_{A}\bigr\rangle$
(and thus of $\Delta\bigl\langle H_{B}\bigr\rangle$) as predicted
by Eq.~(\ref{eq:10}). In particular, the maximization over the set
of SEC unitaries yields 
\begin{align}
\textrm{max}_{\{U\}}\Delta\bigl\langle H_{\gamma}\bigr\rangle & =\textrm{max}_{\{r,\phi\}}\Delta\bigl\langle H_{\gamma}\bigr\rangle\nonumber \\
 & =\frac{\hbar\omega c_{x}}{2}=\frac{\hbar\omega c_{y}}{2},\label{eq:26}
\end{align}
where Eqs. ~(\ref{eq:10}) and (\ref{eq:19}) have been used. To
characterize the behavior of $\textrm{max}_{\{U\}}\Delta\bigl\langle H_{\gamma}\bigr\rangle$
with respect to the entanglement of Bell diagonal states, we shall
obtain the lines of constant concurrence on the region of interest. 

The concurrence $C(\rho)$ is an entanglement measure (monotonically
related to the entanglement of formation) defined as $C(\rho)\equiv\textrm{max}(0,\eta_{4}-\eta_{3}-\eta_{2}-\eta_{1})$,
where $\{\eta_{i}\}$ are the eigenvalues of the hermitean matrix
$R=\sqrt{\sqrt{\rho}(\sigma_{y}\otimes\sigma_{y})\rho^{\ast}(\sigma_{y}\otimes\sigma_{y})\sqrt{\rho}}$,
set in increasing order: $\eta_{i+1}\geq\eta_{i}$. $\sigma_{y}=-i|0\rangle\langle1|+i|1\rangle\langle0|$
is a Pauli matrix and $\rho^{\ast}$ is the state defined through
complex conjugation of $\rho$ in the standard (computational) basis:
$\rho_{i,j}^{\ast}=(\rho_{i,j})^{\ast}$. Since we choose $\alpha_{0,1}^{(1)}$
real, ${\color{black}\rho^{\ast}=\rho}$. Moreover, Bell-diagonal
states are invariant under the application of $\sigma_{y}\otimes\sigma_{y}$.
This implies that $R=\sqrt{\sqrt{\rho}\rho\sqrt{\rho}}=\rho$, which
allows us to compute the concurrence in terms of the eigenvalues of
$\rho$. These eigenvalues are related to the parameters $c_{i}$
by the equation 

\begin{equation}
\lambda_{ab}=\frac{1}{4}\left(1+(-1)^{a}c_{x}-(-1)^{a+b}c_{y}+(-1)^{b}c_{z}\right),\label{eq:21}
\end{equation}
where $\lambda_{ab}$ is the eigenvalue associated to the Bell state
$|\psi_{ab}\rangle=\frac{1}{\sqrt{2}}\left(|0,b\rangle+(-1)^{a}|1,1\oplus b\rangle\right)$,
$a,b=0,1$.\textcolor{red}{{} }Taking into account Eqs. ~(\ref{eq:19})
and ~(\ref{eq:20}), we obtain 
\begin{align}
\lambda_{00} & =\lambda_{10}=\frac{1}{4}\left(1+c_{z}\right)=\frac{1}{2}-p_{0,1},\label{eq:22}\\
\lambda_{11} & =\frac{1}{4}\left(1-2c_{x}-c_{z}\right)=-\textrm{Re}(\alpha_{0,1}^{(1)})+p_{0,1},\label{eq:23}\\
\lambda_{01} & =\frac{1}{4}\left(1+2c_{x}-c_{z}\right)=\textrm{Re}(\alpha_{0,1}^{(1)})+p_{0,1}.\label{eq:24}
\end{align}

For the states we are interested in ($c_{x}\geq0$), Eqs. (\ref{eq:23})
and (\ref{eq:24}) imply that $\lambda_{01}\geq\lambda_{11}$. This
leaves us with the three possible chains of inequalities $\lambda_{00}=\lambda_{10}\geq\lambda_{01}\geq\lambda_{11}$,
$\lambda_{01}\geq\lambda_{00}=\lambda_{10}\geq\lambda_{11}$, and
$\lambda_{01}\geq\lambda_{11}\geq\lambda_{00}=\lambda_{10}$. In the
first case, $\textrm{max}\{\lambda_{ab}\}=\lambda_{00}=\lambda_{10}$
and $C(\rho)=\textrm{max}(0,-\lambda_{01}-\lambda_{11})=0$. For the
remaining possibilities the concurrence yields $C(\rho)=\textrm{max}(0,\lambda_{01}-\lambda_{11}-2\lambda_{00})=\textrm{max}(0,c_{x}-(1+c_{z})/2)$.
Accordingly, $C(\rho)$ is non null for states such that $c_{x}-(1+c_{z})/2>0$.
The lines where $C(\rho)=C$ is constant correspond to equations of
the form $C=c_{x}-(1+c_{z})/2$. In this way, we obtain the following
expression for $c_{z}$ in terms of $c_{x}$:
\begin{equation}
c_{z}=2(c_{x}-C)-1.\label{eq:25}
\end{equation}
This equation represents straight lines that vary in their intersection
with the $c_{z}$ axis, and have constant slope equal to 2. For $C=0$,
$c_{z}=2c_{x}-1$ determines the boundary between entangled and separable
states in Fig. 1(b) of the main text. As $C$ increases, the lines
move towards the bottom left vertice of the entangled triangle, along
the constant direction $\hat{C}$ (see Fig. 1(b)). 

The behavior of $\textrm{max}_{\{U\}}\Delta\bigl\langle H_{\gamma}\bigr\rangle$
with respect to $C$ can now be deduced by using the definition of
directional derivative. Employing Eq.~(\ref{eq:26}), we can calculate
the projection of the gradient of $\textrm{max}_{\{U\}}\Delta\bigl\langle H_{\gamma}\bigr\rangle$
onto the plane determined by Eqs. (\ref{eq:19}) and (\ref{eq:20}).
To this end we just have to compute the components of the gradient
on such a plane. Denoting this projection as $\nabla\left(\textrm{max}_{\{U\}}\Delta\bigl\langle H_{\gamma}\bigr\rangle\right)$,
we obtain:
\begin{equation}
\nabla\left(\textrm{max}_{\{U\}}\Delta\bigl\langle H_{\gamma}\bigr\rangle\right)=\frac{\hbar\omega}{2}(1,1,0)=\frac{\hbar\omega}{\sqrt{2}}\hat{\tau},\label{eq:27}
\end{equation}
where $\hat{\tau}$ is the unit vector with coordinates $(c_{x},c_{y},c_{z})=1/\sqrt{2}(1,1,0)$.
The partial derivative of $\textrm{max}_{\{U\}}\Delta\bigl\langle H_{\gamma}\bigr\rangle$
with respect to the concurrence, $\frac{\partial}{\partial C}\textrm{max}_{\{U\}}\Delta\bigl\langle H_{\gamma}\bigr\rangle$,
is proportional to the directional derivate of $\textrm{max}_{\{U\}}\Delta\bigl\langle H_{\gamma}\bigr\rangle$
along the direction $\hat{C}$, $\hat{C}\cdot\nabla\left(\textrm{max}_{\{U\}}\Delta\bigl\langle H_{\gamma}\bigr\rangle\right)$.
The proportionality factor is the inverse of the derivative of $C$
along such a direction and is positive by definition. Therefore, the
sign of $\frac{\partial}{\partial C}\textrm{max}_{\{U\}}\Delta\bigl\langle H_{\gamma}\bigr\rangle$
is the same of the scalar product $\hat{C}\cdot\hat{\tau}$. Keeping
in mind the geometry of the ``level surfaces'' (within the plane
of interest) for the concurrence, expressed by Eq.~(\ref{eq:25}),
the direction of the corresponding projected gradient is $\hat{C}=1/\sqrt{3}(1,1,-1)$.
In this way, $\hat{C}\cdot\hat{\tau}=\sqrt{2/3}$, and we have a monotic
increment of the maximum energy transfer with respect to $C$: $\frac{\partial}{\partial C}\textrm{max}_{\{U\}}\Delta\bigl\langle H_{\gamma}\bigr\rangle>0$.

Finally, we explicitly compute the value of $\textrm{max}_{\{U\}}\Delta\bigl\langle H_{\gamma}\bigr\rangle$
for states of maximum coherence $\alpha_{0,1}^{(1)}=\sqrt{p_{0,1}p_{1,0}}=p_{0,1}$
(cf. Eq.~(\ref{eq:16})). According to Eqs. ~(\ref{eq:19}) and
~(\ref{eq:20}), $c_{x}=2p_{0,1}$ and $c_{z}=1-4p_{0,1}$ for these
states. In this way, Eq.~(\ref{eq:25}) yields $C=c_{x}-(1+c_{z})/2=4p_{0,1}-1.$
Moreover, $\textrm{max}_{\{U\}}\Delta\bigl\langle H_{\gamma}\bigr\rangle=\hbar\omega p_{0,1}$,
using Eq.~(\ref{eq:26}). This result coincides with Eq.~(9) of
the main text, obtained for two-qubit states that are not necessarily
Bell diagonal, by also optimizing $\Delta\bigl\langle H_{\gamma}\bigr\rangle$
with respect to $\alpha_{0,1}^{(1)}$. Therefore, we arrive at the
expression 
\begin{equation}
\frac{\textrm{max}_{\{U;\alpha_{0,1}^{(1)}\}}\Delta\bigl\langle H_{\gamma}\bigr\rangle}{\hbar\omega}=\frac{\Delta\bigl\langle H_{\gamma}\bigr\rangle_{\textrm{max}}}{\hbar\omega}=\frac{1+C}{4},\label{eq:28}
\end{equation}
which constitutes a simple relation between $\Delta\bigl\langle H_{\gamma}\bigr\rangle_{\textrm{max}}$
and the concurrence. It is worth remarking that Eq.~(\ref{eq:28})
does not mean that the optimal energy transfer only depends on the
concurrence. Since it has been computed over a unidimensional region
(see the red line in Fig. 1(b) of the main text), a dependence with
a single variable is expected. However, we can at least be sure that
$\frac{\partial}{\partial C}\textrm{max}_{\{U\}}\Delta\bigl\langle H_{\gamma}\bigr\rangle>0$,
as was concluded by the previous analysis. 

\begin{thebibliography}{10}
\bibitem{50-comment about SEC U} If $[H_{A}+H_{B},H_{I}]=0$,\textcolor{red}{~}$U=\textrm{exp}\left(-i(H_{A}+H_{B})t\right)\textrm{exp}\left(-iH_{I}t\right)$\textcolor{red}{~}(where
$\hbar=1$ for simplicity). Therefore,\textcolor{red}{~}$[U,H_{A}+H_{B}]=\textrm{exp}\left(-i(H_{A}+H_{B})t\right)[\textrm{exp}(-iH_{I}t),H_{A}+H_{B}]=0$.
\bibitem{52-comment-about-coh} The coefficients $\alpha_{i_{E},j_{E'}}$
must be properly chosen, in such a way that $\rho$ represents a physical
state.
\bibitem{55-generalization of theorems to B} Theorem 3 is explicitly
formulated for both systems, $\gamma=A$ and $\gamma=B$. The extension
of Theorems 1 and 2 to system $B$ is immediate, by simply interchanging
the labels $A$ and $B$. Such a generalization is possible because
both $H_{A}$ and $H_{B}$ are non degenerate Hamiltonians.
\bibitem{58-comment-quantumcorr} Consider a general (bipartite) classically
correlated state, $\rho=\sum_{i}p_{i}|\psi_{i}\rangle_{A}\langle\psi_{i}|\otimes|\psi_{i}\rangle_{B}\langle\psi_{i}|$,
where $\{|\psi_{i}\rangle_{\gamma}\}$ is an orthonormal basis for
system $\gamma$ and $\sum_{i}p_{i}=1$. If $\rho^{\gamma}=\sum_{i}p_{i}|\psi_{i}\rangle_{\gamma}\langle\psi_{i}|$
is diagonal in $H_{\gamma}$, for $\gamma=A,B$, then $\rho=\rho_{Diag}$
is diagonal in the eigenbasis of $H_{A}+H_{B}$.
\end{thebibliography}
\end{document}